\documentclass[aps,twocolumn,superscriptaddress,nofootinbib]{revtex4}
\usepackage{exscale}
\usepackage{graphicx}
\usepackage{amsmath}
\usepackage{latexsym}
\usepackage{amsfonts}
\usepackage{amssymb}

\begin{document}

\title{Double barrier potentials for matter-wave gap solitons}
\author{V. Ahufinger}
\affiliation{ICREA and Grup d'\`Optica, Departament de F\'isica, Universitat Aut\`onoma
de Barcelona, E-08193 Bellaterra, Barcelona, Spain.}
\author{B. A. Malomed}
\affiliation{Department of Physical Electronics, School of Electrical Engineering,
Faculty of Engineering, Tel Aviv University, Tel Aviv 69979, Israel}
\author{G. Birkl}
\affiliation{Institut f\"ur Angewandte Physik, Technische Universit\"at Darmstadt, 64289 Darmstadt, Germany}
\author{R. Corbal\'an}
\affiliation{Grup d'\`Optica, Departament de F\'isica, Universitat Aut\`onoma de
Barcelona, E-08193 Bellaterra, Barcelona, Spain.}
\author{A. Sanpera}
\affiliation{ICREA and Grup de F\'isica Te\`orica, Departament de F\'isica, Universitat
Aut\`onoma de Barcelona, E-08193 Bellaterra, Barcelona, Spain.}

\begin{abstract}
We investigate collisions of solitons of the gap type, supported by a
lattice potential in repulsive Bose-Einstein condensates, with an effective
double-barrier potential that resembles a Fabry-Perot cavity. We identify
conditions under which the trapping of the entire incident soliton in the
cavity is possible. Collisions of the incident soliton with an earlier
trapped one are considered too. In the latter case, many outcomes of
the collisions are identified, including merging, release of the trapped
soliton with or without being replaced by the incoming one, and trapping
of both solitons.
\end{abstract}

\maketitle

\section{Introduction}

\label{sec:1}

Bright matter-wave solitons \cite{bright,Oberthaler} provide an exceptional 
testbed for studying quantum mechanics above the single-atom level. For
the observation of quantum phenomena, coherence is a crucial requirement,
and matter-wave solitons, which propagate without dispersion, may render
this observation more feasible.

The generation of solitary waves in quasi-one-dimensional (Q1D) attractive
Bose-Einstein condensates (BEC) \cite{bright} may be regarded as a tuned
equilibrium between the dispersive effects that tend to spread
the atomic wave function, and the nonlinear attractive interactions which
oppose the spreading by providing an effective self-focusing of the
matter waves. The consequence of such an equilibrium is a stable mesoscopic
atomic wave packet propagating without dispersion.

Repulsive condensates in Q1D geometries can support matter-wave solitons of
the gap type \cite{Oberthaler} if they are loaded in an optical lattice
(OL), which gives rise to the bandgap structure, and are placed at the edge
of the first Brillouin zone, where there is a gap between the first and
second bands \cite{gap-soliton}. Under these conditions, the soliton
exhibits an effective \textit{negative mass}, which permits to balance the
dispersion and the nonlinear interactions, even if they are repulsive.
Various effects generated by the negative effective mass were considered in
Refs. \cite{Oberthaler,malomed1,nosaltres1,nosaltres3}.

Stability conditions for gap solitons (GSs) impose severe restrictions on
their interactions with a potential well or barrier corresponding to a local
modification of the periodic structure. Specifically, the requirement of the
stability allows only for perfect transmission or perfect reflection, but
not partial transmission and reflection that plane waves
display \cite{cohen}. In other words, GSs cannot split through the interaction with
linear defects, and behave like particles exhibiting mesoscopic quantum
features \cite{nosaltres3}, including the quantum reflection of the entire
soliton -- an effect that has been also reported recently for matter-wave
solitons in the self-attractive BEC case \cite{brand}. Although quantum
reflection of ultracold atoms from a solid surface has been reported too
\cite{qrefl}, the limit of the complete ($100\%$) reflection, predicted for
the solitons, is not achievable in that case. Interactions of matter-wave 
solitons with nonlinear traps and barriers produced by spatial variations 
of the scattering length have also been recently addressed \cite{nonlinear}.

In this work we aim to explore such quantum features in the case of the
interaction of a matter-wave soliton of the gap type with a double-barrier
potential resembling a Fabry-Perot cavity. In particular, we demonstrate
that it is possible to trap a soliton in the cavity formed by the two
potential barriers (a similar effect for optical solitons in fiber Bragg
gratings was predicted in Ref. \cite{Peter}). In the context of BEC, the
propagation through a double-barrier potential acting as a Fabry-Perot
interferometer for matter-waves leads to bistability of the transmitted flux
and resonant transport \cite{fabry-perot}.

The paper is organized as follows. In Sec.~\ref{sec:2}, we formulate the
physical model which includes the effective cavity for the GS. 
Sec.~\ref{sec:3} is devoted to the study of the conditions for the trapping of the
entire soliton in the cavity. In Sec.~\ref{sec:4}, the collision of a
second GS with one trapped in the cavity is studied (this interaction also
bears some similarity of collisions between free and defect-trapped optical
solitons in fiber gratings \cite{Peter2}). The paper is concluded in 
Sec.~\ref{sec:5}.

\section{The physical model}

\label{sec:2}

The dynamics of bright GSs created in a Q1D geometry at zero temperature may
be accurately described by the one-dimensional (1D) Gross-Pitaevskii equation
(GPE):
\begin{equation}
i\hbar \frac{d\psi }{dt}=\left[ -\frac{\hbar ^{2}}{2m}\triangle +V\left(
x\right) +g|\psi |^{2}\right] \psi ,  \label{GPE}
\end{equation}%
where the effective nonlinearity is $g\equiv 2\hbar a_{s}\omega _{t}$, with 
$a_{s}$ the $s$-wave scattering length and $\omega _{t}$ the transverse
trapping frequency. The effective axial potential,
\begin{equation}
V\left( x\right) =(1/2)m{\omega _{x}^{2}}x^{2}+V_{0}\sin ^{2}(\pi x/d),
\end{equation}%
includes the parabolic trap with corresponding frequency $\omega _{x}$ and the 
OL with spatial period $d$ and depth $V_{0}$.

First, we briefly summarize the numerical procedure used to generate GSs, as
per Ref. \cite{nosaltres1}. The ground state is found for a $^{87}$Rb
condensate ($a_{s}=5.8$ nm) formed by $N=500$ atoms trapped magnetically
with transverse and axial frequencies $\omega _{t}=715\times 2\pi $ Hz and 
$\omega _{x}=14\times 2\pi $ Hz, respectively, in the presence of an OL, with
potential depth equal to the OL recoil energy, $V_{0}=E_{r}\equiv \hbar
^{2}k^{2}/2m$, where $k=\pi /d$ is the recoil momentum, and $d=397.5$nm, the 
period. Then, the axial magnetic trap is suddenly turned off and an appropriate 
phase imprinting leads to the inversion of the sign of the
wave function at each second site. As a result, the system
evolves toward a self-sustained staggered soliton with a negative effective
mass. The soliton, which contains approximately $35\%$ of the initial number
of atoms and extends over $\simeq 11$ sites of the OL potential, is
generated at rest with the center at $x=0$. In order to set
it into motion, we instantaneously impart an appropriate momentum to the
soliton, to which it responds by self-adaptation to the new conditions, i.e., 
expunging atoms until a new equilibrium state is reached. With
momentum $p=0.1k\hbar $ lent to the GS, it settles down into a 
state with $27\%$ of the initial number of atoms ($N_{\mathrm{final}}=135$),
total energy $0.92E_{r}$, and the kinetic energy of its motion 
$E_{k}=0.01E_{r}$ \cite{nosaltres3}.

Our aim is to study the interaction of the so generated moving matter-wave
soliton with a Fabry-Perot type potential, formed by two potential barriers
forming a cavity for the soliton, cf. a similar configuration proposed for
optical solitons in fiber gratings \cite{Peter}. After turning off the
magnetic trap, $V(x)$ in Eq. (\ref{GPE}) accounts only for the OL,
which is locally modified around two sites, $x_{m1}$ and $x_{m2}$,
 to generate the double-barrier structure as follows:

\begin{equation}
V\left( x\right) =V_{0}\sin ^{2}(\pi x/d)+V_{\mathrm{mod}}\left( x\right) ,
\label{pot}
\end{equation}

where $V_{\mathrm{mod}}\left( x\right)$ reads:
\begin{equation}
  \left\{ \begin{array}{ll}
       V_{m1}(1-\frac{(x-x_{m1})^2}{2\sigma^2})& \mbox{\small{if $x_{m1}-l/2\leq x\leq x_{m1}+l/2$}};\\
       V_{m2}(1-\frac{(x-x_{m2})^2}{2\sigma^2})& \mbox{\small{if $x_{m2}-l/2\leq x\leq x_{m2}+l/2$}};\\
       0   & \mbox{\small{otherwise}}.\end{array} \right. 
\label{potmod} 
\end{equation}with $\sigma =6d$. Below, we consider the symmetric cavity created by
two identical barriers, with $V_{m1}=V_{m2}=V_{m}<0$. Note that, due to the
negative effective mass of the GS, the barriers actually correspond to a
local \emph{decrease} of the periodic potential. Points $x_{m1,2}$ are fixed
at local minima of the OL potential, and the distance between the barrier centers, 
$A\equiv x_{m2}-x_{m1}$, is assumed large enough to have the actual size of
the cavity, $B=A-l$, much larger than the axial size of the soliton.

\section{The cavity}

\label{sec:3}

\begin{figure}[tbp]
\includegraphics[width=1.0\linewidth]{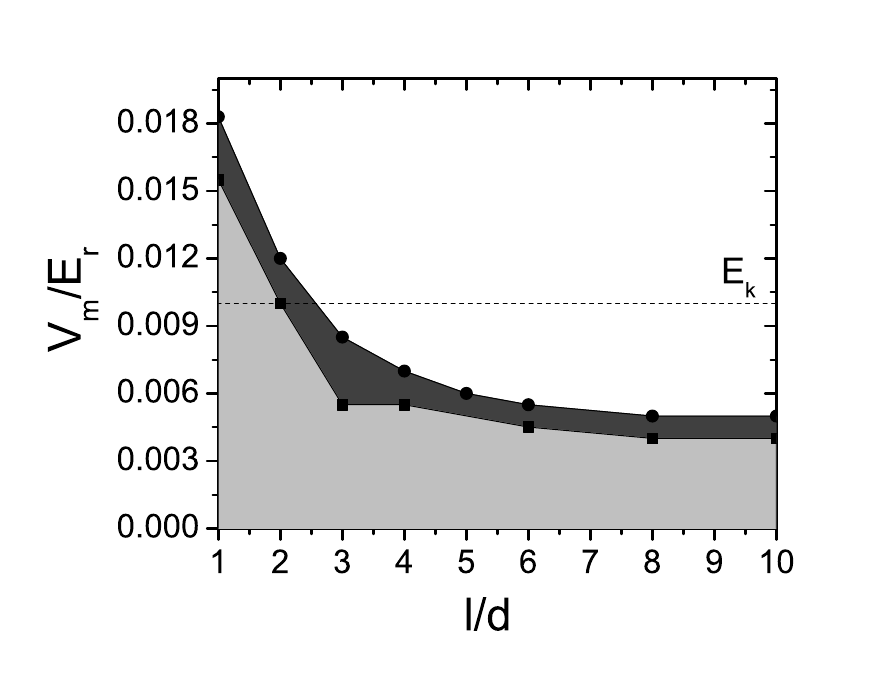}
\caption{Diagram of outcomes of the collisions of a moving lattice soliton
with a cavity formed by two identical barriers: reflection, transmission,
and trapping, in the white, light grey, and dark grey areas, respectively.
The effective size of the cavity is $B=20d$. Parameters $l$ (measured in
units of the lattice period, $d$) and $V_{m}$, measured in units of the
recoil energy, $E_{r}$, are the width and height of the two barriers, respectively. The
horizontal dotted line designates the kinetic energy of the moving soliton.}
\label{diagram}
\end{figure}

\begin{figure}[tbp]
\includegraphics[width=1.0\linewidth]{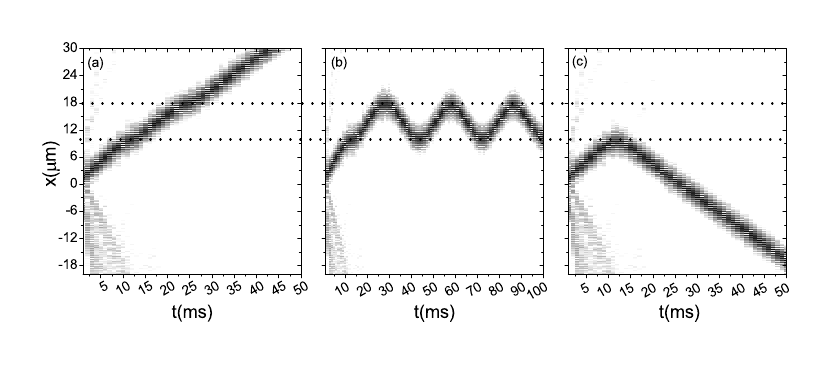}
\caption{Contour plots of the spatiotemporal evolution of the lattice
soliton colliding with the set of two identical barriers centered at 
$x_{m1}=25d$ and $x_{m2}=45d$, each with width $l=2d$ and the following
height: (a) $|V_{m}|=0.008E_{r}$; (b) $|V_{m}|=0.011E_{r}$; 
(c) $|V_{m}|=0.014E_{r}$. Here as well as in the following figures 
the horizontal dotted lines show the position of the centers of both barriers.}
\label{amplitude}
\end{figure}

\begin{figure}[tbp]
\includegraphics[width=1.0\linewidth]{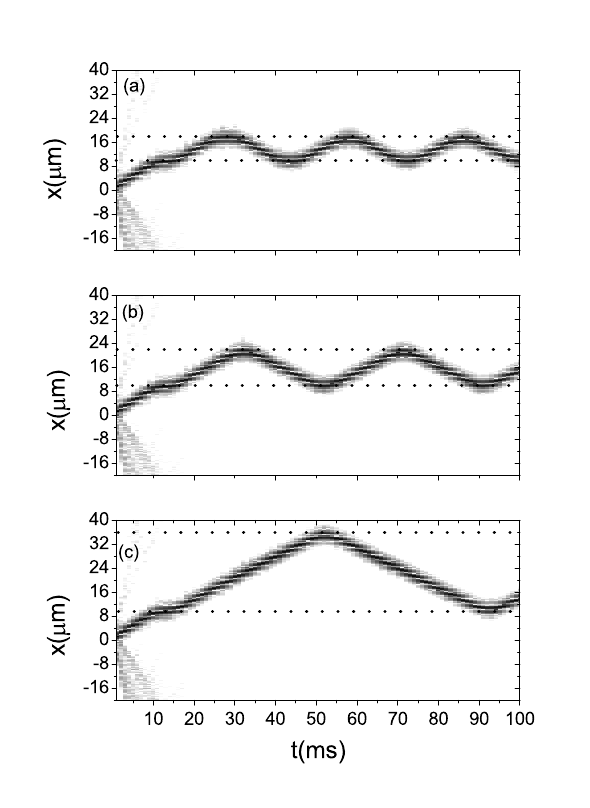}
\caption{Examples of the trapping of the lattice soliton by the pair of
barriers of width $l=2d$ and height $|V_{m}|=0.011E_{r}$. The first barrier
is centered at $x_{m1}=25d$, and the second at: (a) $x_{m2}=45d$, 
(b)$x_{m2}=55d$, and (c) $x_{m2}=90d$. }
\label{distance}
\end{figure}

The interaction of a moving matter-wave GS with a single barrier in the OL
was addressed recently in Ref. \cite{nosaltres3}, where it was shown that
the soliton does not split. For a fixed kinetic energy, there exists an
abrupt transition from complete transmission to complete reflection, as the
height of the barrier increases, for all considered values of the barrier
widths. The border between these two outcomes of the collision of the
soliton with the barrier is shown, in the plane of the barrier's parameters,
height $V_{m}$ and width $l$, by filled circles in Fig. \ref{diagram}.
Complete reflection and transmission occur, respectively, above and below
this border (i.e., the white area in the figure corresponds to the bounce of
the entire soliton from the barrier). In the process of the collision, the soliton
naturally decreases its velocity (even if does not bounce). This slowing down of 
the soliton in the region of the defect is more pronounced as we approach the 
border between the two behaviors.
Addressing the configuration with the second barrier
placed at a certain distance from the first one, we notice that the behavior
of the incident soliton which hits the first barrier remains as described
before, i.e., a sudden transition from perfect transmission to perfect
reflection occurs. Nevertheless, when the height of the barriers is close to
the transition point, and the length of the cavity ($B$) is larger than the
size of the soliton, the soliton which has passed the first barrier gets
trapped in the cavity, in a state of oscillatory motion. Thus, three
scenarios are identified in the interaction of the incident soliton with the
double-barrier structure: complete reflection, complete transmission, and
trapping into the oscillatory state. As said above, the reflection occurs in
the same area of the parameter space (white region in Fig. \ref{diagram}) as
for the single barrier. The complete transmission takes place for values of
the barrier's height and width well inside the transmission region for the
single barrier (the light grey area in Fig. \ref{diagram}), while the
trapping is observed close to the transmission-reflection border for the
single barrier (the dark grey region in Fig. \ref{diagram}). The results
shown in Fig. \ref{diagram} correspond to a fixed cavity size of $B=20d$ and
we have checked that the parameter space for trapping slightly increases with the size of
the cavity. For a fixed width of the barriers, and a large enough distance
between them, the three scenarios follow each other with the increase of the
barriers' height. These scenarios are illustrated, in Fig. \ref{amplitude} by spatiotemporal
trajectories of the soliton hitting the cavity formed by two identical
barriers of width $l=2d$, which are separated by distance $A=20d$, giving 
the cavity enough room to trap the soliton, $B=18d$. The complete
transmission is displayed in panel (a) for $|V_{m}|=0.008E_{r}$, panel (b)
with $|V_{m}|=0.011E_{r}$ shows an example of the trapping, and the bounce
(complete reflection) is observed in panel (c), for $|V_{m}|=0.014E_{r}$.
Figure \ref{amplitude}(b) clearly demonstrate that the trapped soliton
performs periodic oscillations in the cavity without any visible loss of
atoms. The period of the oscillations, for given parameters of the barriers,
can be modified by changing the size of the cavity, as shown in Fig. \ref{distance}. 
Note that the smallest period, limited by the condition that the
size of the cavity must be larger than the soliton's axial size, is $\simeq
30$~ms, in physical units. Faster oscillations were predicted for a lattice
soliton trapped in a potential well \cite{nosaltres3}.
It is relevant to note that, in the above-mentioned case of the trapping of
optical solitons by a cavity in the model of the fiber grating \cite{Peter},
 the maximum capture efficiency (in terms of the soliton's energy) is no
more than $60\%$, contrary to what happens with the matter-wave GSs, which
may be trapped \emph{entirely}, without losses.

\section{Collisions between free and trapped solitons}

\label{sec:4}
\begin{figure}[tbp]
\includegraphics[width=1.0\linewidth]{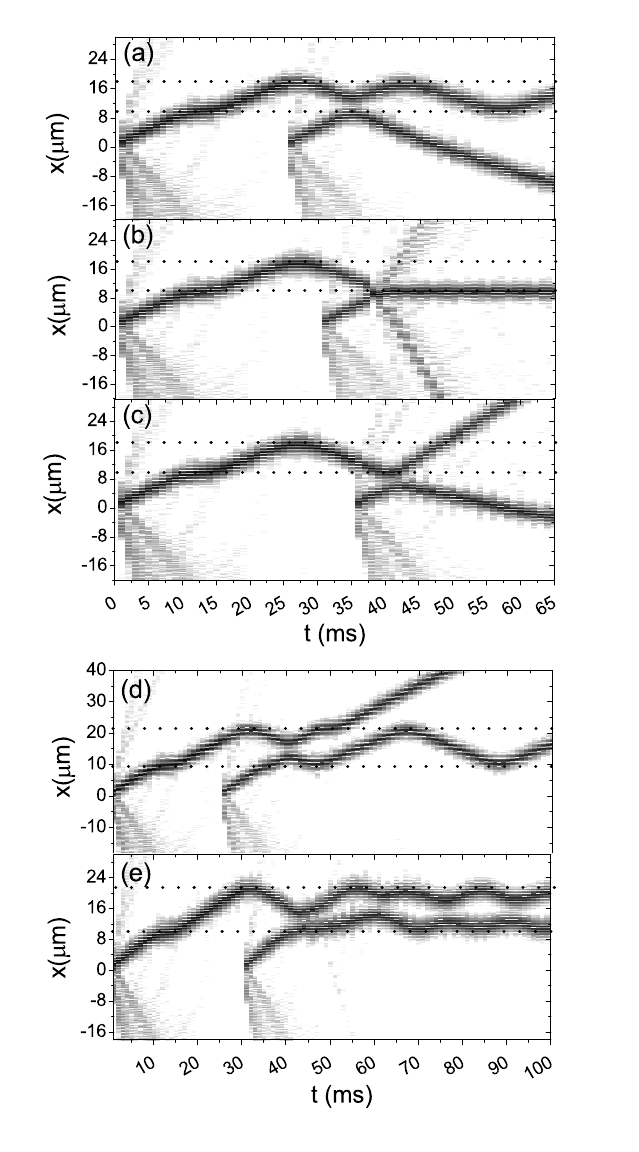}
\caption{Different outcomes of the collision between an incident GS with an 
identical one previously trapped. The
barriers of width $l=2d$ and height $|V_{m}|=0.011E_{r}$ are separated by
distance $A=20d$, in (a), (b) and (c), or $A=30d$, in (d) and (e). The
second soliton is send after the first one with time delay $\Delta t=20$~ms
(a), $\Delta t=30$~ms (b), $\Delta t=35$~ms (c), $\Delta t=25$~ms (d), and 
$\Delta t=30$~ms (e).}
\label{two}
\end{figure}

The next natural step in the analysis is to consider a collision between an
incident soliton with the cavity already occupied by an (identical) earlier
trapped GS. Different outcomes of the collisions are observed, depending on
time delay $\Delta t$ between the two solitons. Figure \ref{two} displays
three cases for the cavity formed by two barriers of width $l=2d$ and height
$|V_{m}|=0.011E_{r}$, separated by a distance of $A=20d$, corresponding to 
(a) $\Delta t=20$~ms, (b) $\Delta t=30$~ms, and (c) $\Delta t=35$~ms. In (a), the
incident soliton bounces back, while the trapped one performs oscillations
in the cavity; in (b), the two solitons \emph{merge} into a single one,
and in (c), the incident soliton bounces back, kicking out the trapped one
in the forward direction. Two more cases are shown in Fig. \ref{two} for the
same parameters of the barriers, but for a different separation between them, 
$A=30d$, and time delays $t=25$~ms in (d), and $t=30$~ms (e). In this
case, the cavity has enough room to trap the two solitons, which gives
rise to new collision scenarios: in (d), the incoming soliton gets
trapped by kicking out the previously trapped one (``recharge"),
while in (e) \emph{both} solitons get trapped in the cavity,
oscillating in counter-phase. These sundry dynamical behaviors
suggest new experimental possibilities for the control and
manipulation of matter-wave GSs.
\vspace{0.5cm}
\section{Conclusions}

\label{sec:5}
In this work, we have studied the interactions of bright matter-wave
solitons of the gap type, with negative effective mass, which are
supported by the interplay of the OL (optical lattice) and repulsive
nonlinearity in BEC, with a cavity formed by two far separated identical
local potential barriers. We have shown that there exists a parameter region
in which the incident soliton is trapped by the cavity into the shuttle
state. This region can be found for all the values of the barriers' width,
provided that their height corresponds to the transmission of the soliton by
a single barrier, but close to the reflection-transmission border.

The interaction of a second soliton which hits the cavity already
occupied by a trapped oscillating soliton has been considered too.
In that case, a number of different collision scenarios can be
identified, depending on the time delay between the launch of the
two solitons. Particularly interesting outcomes are the merging of
the two solitons at the position of the first barrier, bounce of the
second soliton kicking out the trapped one, the ``recharge",
i.e., release of the originally trapped soliton which is replaced by
the incident one, and trapping of both solitons into the state of
shuttle oscillations in the cavity, with a phase shift of $\pi $
between them.

We acknowledge support from the Spanish Ministerio de Ciencia y 
Tecnolog\'{\i}a (FIS2005-01369, FIS2005-01497, Consolider Ingenio 2010
CSD2006-00019), from the scientific exchange programme Spain-Germany 
(MEC under contract HA2005-0002 and DAAD under contract D/05/25694), 
from the Catalan Government (SGR2005-00358), from the European 
Commission (Integrated Project SCALA), from the Israel Science 
Foundation (Center-of-Excellence grant No. 8006/03) and
German-Israel Foundation (grant No. 149/2006).

\end{document}